\documentstyle[seceq,epsf]{ptptex}

\title{
Spin dynamics simulations -- a powerful method for the study of critical dynamics
}

\author{
D. P. {\sc Landau},
Alex {\sc Bunker}\footnote{Present address: Max Planck Institute
 for Polymer Research, Ackermann Weg 10, Mainz, Germany D-55021-3148},
Hans Gerd {\sc Evertz}\footnote{Present address: Inst. f. Theor. Physik, 
TU Graz, 8010 Graz, Austria},
M. {\sc Krech},
and Shan-Ho {\sc Tsai}
}

\inst{
Center for Simulational Physics, University of Georgia, Athens, GA 30602, USA}

\recdate{
\today
}

\abst{

Spin-dynamics techniques can now be used to study the deterministic time-dependent
behavior of magnetic systems containing over $10^5$ spins with quite good accuracy. 
This approach will be introduced, including the theoretical foundations of the methods 
of analysis. Then newly developed, improved techniques based upon Suzuki-Trotter 
decomposition methods will be described. The current ``state-of-the-art'' will be 
evaluated with specific examples drawn from data on simple magnetic models. The 
examination of dynamic critical behavior will be highlighted but the extraction of 
information about excitations at low temperatures will be included.

}

\begin{document}

\maketitle

\section{Introduction}

The static behavior of physical systems near continuous phase transitions 
is characterized by a set of static critical exponents, which describe the
critical behavior of thermodynamic quantities such as the specific heat, 
the order parameter, the correlation length, and so on. One can thus define
different universality classes, within which the critical exponents are identical.
The numerical values of the critical exponents depend only on the symmetry of
the order parameter, the dimensionality of the system and the range of interactions,
but not on either the precise form of the model Hamiltonian or the lattice type. 
Likewise, the dynamic critical behavior is describable in terms of a dynamic
critical exponent $z$, which depends on the conservation laws and which, in 
analogy to static critical phenomena, gives rise to different dynamic universality
classes\cite{HohHal77}. Our understanding of static critical behavior is now
mature and has resulted largely from the
investigation of simple model spin systems such as the Ising,
the XY, and the Heisenberg model. These models are equally
valuable for the investigation
of dynamic critical behavior and dynamic scaling. 
Realistic models of magnetic materials can be constructed from these
simple spin models; however, the theoretical analysis of
experimentally accessible quantities, such as the dynamic structure
factor, is usually too demanding for analytical methods. Computer
simulations are
beginning to provide important information about dynamic critical
behavior and material properties of model magnetic systems.
\cite{EvLan96,ChenLan94,BunChenLan96} These simulations use
model Hamiltonians with continuous degrees of freedom represented by a
three-component spin ${\bf S}_j$ with fixed length $|{\bf S}_j| = 1$
for each lattice site $j$. A typical model Hamiltonian is then given
by
\begin{equation}
\label{H}
{\cal H} = -J \sum_{<j,l>} \left(S_j^x S_l^x + S_j^y S_l^y + \lambda
S_j^z S_l^z \right) - D \sum_j \left( S_j^z \right)^2 ,
\end{equation}
where $J$ is the coupling constant, $<j,l>$ denotes a nearest-neighbor
pair of spins, $\lambda$ is an anisotropy parameter, and $D$
determines the strength of a single-site or crystal field anisotropy.
(We use units in which Boltzmann's constant $k_B=1$.)  

The dynamics of the spins are governed 
by the coupled equations of motion \cite{GerLan90}
\begin{equation}
\frac{d}{dt}{\bf S}_j = \frac{\partial {\cal H}}{\partial {\bf S}_j} \times {\bf S}_j ,
\label{eqmot}
\end{equation}
and the time dependence of each spin can be determined from the
integration of these equations, where (hybrid) Monte-Carlo simulations of the
model provides {\em equilibrium} configurations as initial conditions
for Eq.(\ref{eqmot}).

The most important quantity to be extracted from the numerical results is the
dynamic structure factor $S({\bf q},\omega)$ for momentum transfer
${\bf q}$ and frequency transfer $\omega$, which can be measured
in neutron scattering experiments, and is given by
\begin{equation}
S^k({\bf q},\omega)=\sum_{\bf R} e^{i {\bf q}\cdot {\bf R}}
\int_{-\infty}^{+\infty} e^{i\omega t} C^k({\bf R},t) \frac{dt}{\sqrt{2\pi}},
\end{equation}
where ${\bf R} = {\bf r}_j - {\bf r}_l$ (${\bf r}_j$ and ${\bf r}_l$ are lattice vectors),  
$C^k({\bf R},t)$ is the space- and time-displaced correlation
function, with $k=x, y,$ or $z$, defined as
$C^k({\bf R},t) =\langle {S_j}^k(t){S_l}^k(0)\rangle-
\langle {S_j}^k(t)\rangle\langle {S_l}^k(0)\rangle.$

Two practical limitations on spin-dynamics techniques 
are the finite lattice size and the finite evolution time. The finite time 
cutoff can introduce oscillations in $S^k({\bf q},\omega)$, which can be 
smoothed out by convoluting the spin-spin
correlation function with a resolution function in frequency, which is equivalent 
to the energy resolution in neutron-scattering experiments, yielding $\bar S^k({\bf q},\omega)$.
Finite-size scaling theory \cite{ChenLan94,RapLan96} can be used to extract the 
dynamic critical exponent $z$:  the divergence of the correlation length $\xi$ 
is limited by $L$ and the dynamic finite-size relations are given by
\begin{equation}
\frac{\omega\bar {S_L}^k({\bf q},\omega)}{{\bar{\chi}_L}^k({\bf q})}=
G^k(\omega L^z,qL,\delta_{\omega}L^z)
\end{equation}
and 
\begin{equation}
\bar\omega^k_m=L^{-z}\bar{\cal W}^k(qL,\delta_{\omega}L^z),
\label{omegam}
\end{equation}
where ${\bar{\chi}_L}^k({\bf q})$ is the total integrated intensity and 
$\bar\omega^k_m$ is a characteristic frequency, defined as
\begin{equation}
\int_{-\bar\omega^k_m}^{\bar\omega^k_m}\bar {S_L}^k({\bf q},\omega)\frac{d\omega}{2\pi}
=\frac{1}{2}{\bar{\chi}_L}^k({\bf q}).
\end{equation}

To speed up the numerical integration of Eq.(\ref{eqmot}) it is desirable 
to use the largest possible time step; however, with
standard methods the size of the time step is severely limited
by the accuracy within which the {\em conservation laws} of the dynamics
are obeyed. It is evident from Eq.(\ref{eqmot}) that the
total energy is conserved, and if, for example, $D = 0$ and $\lambda = 1$ (isotropic
Heisenberg model) the magnetization ${\bf M} = \sum_j {\bf S}_j$ is also
conserved. For the anisotropic Heisenberg model, i.e.,
$\lambda \neq 1$ or $D \neq 0$ only $M_z$
is conserved. Conservation of spin length and energy
is particularly crucial, because the condition $|{\bf S}_j| = 1$ is a
major part of the definition of the model and the energy of a
configuration determines its statistical weight. It would therefore
also be desirable to devise an algorithm which conserves these two
quantities {\em exactly}.

The remaining sections of this paper are organized as follows. In Section 2
we describe integration methods, including a newly developed technique based
on Suzuki-Trotter decompositions of exponential operators.\cite{FraHuaLei,KreBunLan} 
In Section 3 we
discuss two examples of physical systems, namely the two-dimensional XY model and
the three-dimensional Heisenberg antiferromagnet, and comparison with experiment
and theory. The purpose of these examples is to show just how far the state-of-the-art
has developed in producing and analyzing spin-dynamics data.
In Section 4 we give a brief summary of this paper. 

\section{Integration methods}

\subsection{Predictor-corrector method}

Predictor-corrector methods have been quite effective for the
numerical integration of spin equations of motion; however, in order
to limit truncation errors small time steps $\delta t$ must be used
with at least a 4th-order scheme.  The predictor
step of the scheme used here is the explicit Adams-Bashforth
four-step method \cite{NumAna81} and the
corrector step consists of typically one iteration of the implicit
Adams-Moulton three-step method.\cite{NumAna81}
This predictor-corrector method is very general and is independent of
the special structure of the equations of
motion (see Eq.(\ref{eqmot})). The conservation laws discussed earlier
will only be observed within the accuracy set by the
truncation error of the method. In practice, this limits the time step
to typically\cite{BunChenLan96} $\delta t = 0.01/J$ in $d = 3$,
where the total integration time is typically $600/J$ or less.

\subsection{Suzuki-Trotter decomposition methods}

The motion of a spin may be viewed
as a precession of the spin ${\bf S}$ around an effective axis
$\Omega$ which is itself time dependent.  The lattice
can be decomposed into two sublattices such that a spin on one
sublattice precesses in a local field $\Omega$ of neighbor
spins which are {\em all} located on the other sublattice. For the
Hamiltonian in Eq.(\ref{H}) there are only two such sublattices
if the underlying lattice is simple cubic.

To illustrate the method, we consider first the $D=0$ case.
The basic idea of the algorithm is to rotate a spin about its local field
$\Omega$ by an angle $\alpha = |\Omega|\delta t$, rather than directly
integrate Eq.(\ref{eqmot}). This procedure
guarantees the conservation of the spin length $|{\bf S}|$
and energy to within machine accuracy.
 Denoting the two sublattices by $\cal A$ and $\cal B$,
respectively, we can express the local fields acting on the
spins on sublattice $\cal A$ and $\cal B$ as 
$\Omega_{\cal A}[\{{\bf S}\}]$ and $\Omega_{\cal B}[\{{\bf S}\}]$, 
respectively. 
In a more symbolic way, we denote $y$ as a complete spin configuration, which
is decomposed into two sublattice components $y_{\cal A}$
and $y_{\cal B}$, i.e. $y = (y_{\cal A},y_{\cal B})$, and we denote 
by matrices $A$ and $B$ the
generators of the rotation of the spin configuration $y_{\cal A}$ on
sublattice $\cal A$ at fixed $y_{\cal B}$ and of the spin
configuration $y_{\cal B}$ on sublattice $\cal B$ at fixed $y_{\cal
A}$, respectively. The update of the configuration $y$ from time $t$
to $t + \delta t$ is then given by an exponential (matrix) operator
\begin{equation}
\label{eAB}
y(t+\delta t) = e^{(A + B)\delta t} y(t) .
\end{equation}
The exponential operator in Eq.(\ref{eAB}) rotates each spin
of the configuration and it has no simple explicit form, because the
rotation axis for each spin depends on the configuration itself; however, 
the set of equations of motion for spins on one sublattice 
reduces to a {\em linear} system of
differential equations if the spins on the other sublattice are kept
fixed and the
operators $e^{A \delta t}$ and $e^{B \delta t}$ which rotate $y_{\cal
A}$ at fixed $y_{\cal B}$ and $y_{\cal B}$ at fixed $y_{\cal A}$,
respectively, {\em do} have a simple explicit form.\cite{KreBunLan} 
Thus an {\em alternating} update scheme may be used, i.e., we
rotate $y_{\cal A}$ at fixed $y_{\cal B}$ and vice-versa.
The alternating update scheme 
amounts to the replacement $e^{(A + B) \delta t} \to e^{A \delta t}
e^{B \delta t}$ in Eq.(\ref{eAB}), which is only correct \cite{SuzUme93}
up to order $(\delta t)^2$. The magnetization will
therefore only be conserved up to terms of the order $\delta t$ (global
truncation error). To decrease truncation error and thus to improve the 
conservation, one can employ $m$th-order
Suzuki-Trotter decompositions of the exponential operator in
Eq.(\ref{eAB}), namely\cite{SuzUme93}
\begin{equation}
\label{epABA}
e^{(A + B) \delta t} = \prod_{i=1}^u e^{p_i A \delta t / 2} e^{p_i B
\delta t} e^{p_i A \delta t / 2} + {\cal O}(\delta t^{m+1})
\end{equation}
where $u=1$ for 2nd-order, $u=5$ for 4th-order, and $u=15$ for 8th-order 
and the parameters $p_i$ are given by Suzuki and Umeno.\cite{SuzUme93} 

The additional computational effort needed to evaluate higher order
expressions can be compensated to some extent
by using larger time steps. 
The inclusion of next-nearest neighbor bilinear interactions on a simple cubic lattice
can be treated within the above
framework if the lattice is decomposed into four sublattices.

This approach can also be extended to the case $D \neq 0$, but in
contrast to the isotropic case, the equation of
motion for each individual spin on each sublattice is {\em nonlinear}.
In practice, the best form of solution is via iterative numerical methods.  
In order to perform a rotation operation in
analogy to the isotropic case we identify an effective rotation
axis
$\widetilde{\Omega}_j = \Omega_j - D \left(0,0,S_j^z(t) +
S_j^z(t+\delta t) \right)$, such that 
the condition for energy conservation is rewritten in the form
$\widetilde{\Omega}_j \cdot ({\bf S}_j(t + \delta t) - {\bf S}_j(t)) =
0$.
Since the rotation requires knowledge of $S_j^z$ at the future time $t +
\delta t$, we use an
iterative procedure starting from the initial value $S_j^z(t+\delta t)
= S_j^z(t) + (\Omega_j\times S_j(t))^z \delta t$ 
and performing several updates according
to the decompositions given by Eq.(\ref{epABA}) in order to improve 
energy conservation.
Both the degree of conservation and the execution time depend to some
extent on the number of iterations used. 

\begin{figure}[htb]
\vspace{-4cm}
\epsfxsize=3.5in
\centerline{\epsfbox{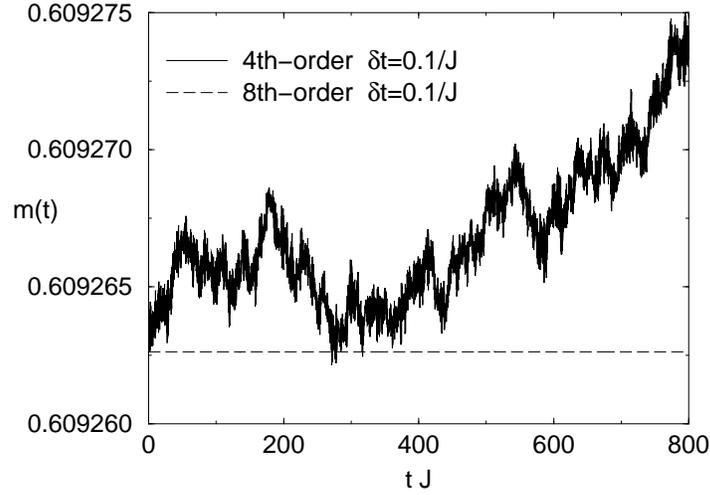}}
\vspace{-1.8cm}
\caption{Magnetization $m(t) = |{\bf M}(t)|/L^3$ per spin for different order
decomposition schemes for $D=0$ and time step $\delta t = 0.1/J$: 
(solid line) 4th-order scheme; (dashed line) 8th-order method.}
\label{szt48}
\end{figure}

\begin{figure}[htb]
\vspace{-2cm}
\epsfxsize=3.9in
\centerline{\epsfbox{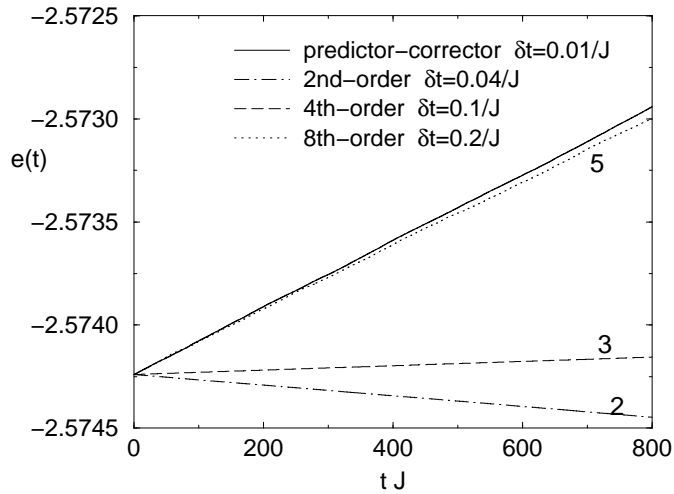}}
\vspace{-2.1cm}
\caption{Energy $e(t)=E(t)/(JL^3)$ per spin for different order
decomposition schemes for $D=J$: (solid line) predictor-corrector method;
(dot-dashed line) 2nd-order scheme; (dashed line) 4th-order scheme; (dotted
line) 8th-order method. The number of iterations performed are marked next
to each line.}
\label{DJcomp}
\end{figure}

For a quantitative analysis of the integration methods outlined
above we restrict ourselves to the Hamiltonian 
given by Eq.(\ref{H}) for $\lambda = 1$ in $d = 3$ and $J>0$. The underlying
lattice is simple cubic with $L = 10$ lattice sites in each direction
and periodic boundary conditions.
In order to compare the different integration methods we 
investigate the accuracy within which the conservation laws are
fulfilled. The initial configuration is a well
equilibrated one from a Monte-Carlo simulation for $\lambda = 1$
at a temperature $T = 0.8T_c$ for $D=0$ and $D=J$, where $T_c$ refers
to the critical temperature of the isotropic model $(D = 0)$. 
Fig.\ref{szt48} shows the magnetization conservation for the 4th- and 8th-order 
decomposition methods, both with $\delta t=0.1/J$, for $D=0$ and Fig.\ref{DJcomp} 
shows the energy conservation for different methods for $D=J$. 
In the latter
case the iterative nature of all four methods gives rise to a basically linear
energy change. A single integration step using the 2nd-, the 4th- and the 8th-order
scheme is respectively about 2 times faster, 2.5 and 9 times slower
than the predictor-corrector method; the speedup of the decomposition methods comes from
the much larger $\delta t$ that can be used. For $D=0$ it is still feasible to use the 4th-order
method with $\delta t=0.2/J$, which corresponds to an eightfold speedup as compared to the 
predictor-corrector method. The 8th-order method improves the conservation significantly 
but at the cost of greatly increased execution time.

\section{Examples of physical systems}

\subsection{Two-dimensional XY model}

The two-dimensional XY model can be described by the Hamiltonian in Eq.(\ref{H})
with $\lambda=0$ and $D=0$. 
At the critical temperature\cite{EvLan96}  $T_{KT}=0.700(5)J$, the model undergoes an
unusual phase transition to a state with bound, topological excitations (vortex pairs), 
and the static properties are consistent with the
predictions of the Kosterlitz-Thouless theory.\cite{KosTho73} 
For $T\le T_{KT}$ the model is critical,
i.e. the correlation length is infinite, but there is no long-range order, and the
spin-spin correlation function decays algebraically with distance, with an exponent 
$\eta$ that varies with temperature. 

In our studies of the dynamics of this model\cite{EvLan96}, we used $L\times L$ lattices with 
periodic boundary conditions for $16\le L\le 192$ and several values of $T$.
The equations of motion (see Eq.(\ref{eqmot})) were integrated using a 4th-order
predictor-corrector method, with a time step of $\delta t=0.01/J$, to a maximum time
$t_{max}=400/J$. Between 500 and 1200 equilibrium configurations were used for each
lattice size and temperature. We were limited to the [10] reciprocal lattice 
direction, i.e. ${\bf q}=(q,0)$ and $(0,q)$.

\begin{figure}
\vspace{-4.5cm}
\epsfxsize=3.9in
\centerline{\epsfbox{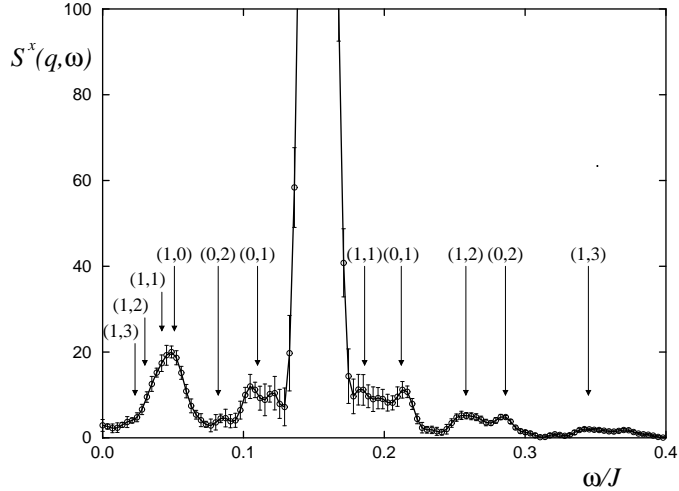}}
\vspace{-2.2cm}
\caption{Low-intensity structure in $S^x(q,\omega)$ for T=0.6J, L=192 and $q=\pi/32$.
Vertical arrows show the location of two-spin-wave peaks formed by spin waves of small 
momentum $q<4(2\pi/L)$.}
\label{F5e}
\end{figure}

\begin{figure}
\vspace{-4.cm}
\epsfxsize=3.8in
\centerline{\epsfbox{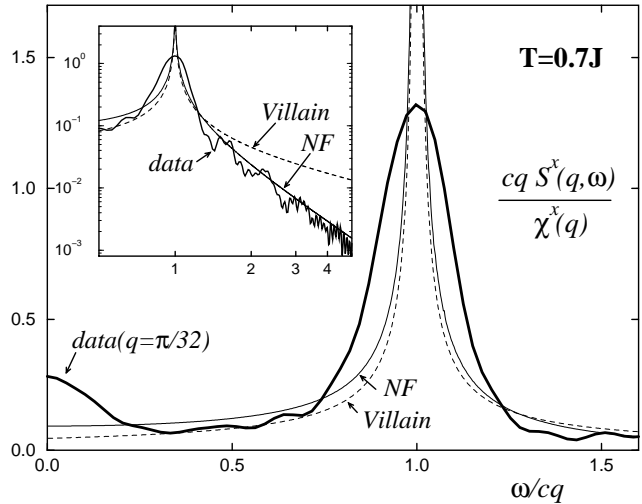}}
\vspace{-2.2cm}
\caption{Comparison of the line shape of $S^x(q,\omega)$ with theoretical predictions.
Data are at $T=T_{KT}$, $L=128$, and $q=\pi/32$ (thick line), normalized as 
$cqS^x(q,\omega)/\chi^x(q)$, where $c$ is the spin-wave velocity. The two thin lines
represent the predictions by Nelson and Fisher\cite{NelFis77} (continuous line) and 
by Villain\cite{Vil74} (dashed line), both with $\eta=0.25$. The inset shows the data and
predictions on a log-log plot that includes large values of $\omega$.}
\label{F13mod}
\end{figure}

For $T\le T_{KT}$, the in-plane component $S^x(q,\omega)$ exhibits very strong and sharp spin-wave
peaks. As T increases, they widen slightly and move 
to lower $\omega$, but remain pronounced even just above $T_{KT}$. For increasing momentum 
they broaden and rapidly lose intensity. Well above $T_{KT}$, the spin-wave peak disappears 
in $S^x(q,\omega)$, as expected, and we observe a large central peak instead. Besides 
the spin-wave peak,  $S^x(q,\omega)$
exhibits a rich low-intensity structure, which we interpret as two-spin-wave processes
(see Fig.\ref{F5e}). Furthermore, $S^x(q,\omega)$ shows a clear central peak, even below 
$T_{KT}$, which becomes very pronounced towards $T_{KT}$. Neither this strong central peak
nor the additional structure are predicted by existing analytical calculations. The 
out-of-plane component $S^z(q,\omega)$ is much weaker than $S^x(q,\omega)$, except for large $q$. 
The very sharp spin-wave peaks at low temperatures allowed us to determine the dispersion
curves with great accuracy. Our estimated value of the dynamic critical exponent is
$z=1.00(4)$, in agreement with the theoretical prediction of $z=1$.
The line shape of $S^x(q,\omega)$ is not well described by either Villain's\cite{Vil74} or 
Nelson and Fisher's\cite{NelFis77} 
prediction (the latter agrees qualitatively with our data only for large $q$) 
(see Fig.\ref{F13mod}). Moreover, these predictions do not describe the additional
structure in $S^x(q,\omega)$, including the central peak.

\subsection{Three-dimensional Heisenberg antiferromagnet and RbMnF$_3$}

RbMnF$_3$ is a good physical realization of an isotropic
three-dimensional Heisenberg antiferromagnet, described by the Hamiltonian
in Eq.(\ref{H}) with $\lambda =1$, $D=0$ and $J<0$. 
Early experimental studies [see references in Ref.\citen{TsaBunLan}] showed that 
the Mn$^{2+}$ ions, with spin $S=5/2$, form a simple cubic lattice structure 
with a nearest-neighbor exchange constant $|J^{exp}|=0.58(6)$ meV and a 
second-neighbor interaction constant of less than $0.04$ meV, both defined 
using the normalization as in Eq.(\ref{H}). 
Magnetic ordering with antiferromagnetic alignment of spins occurs below
the critical temperature $T_c = 83K$. The magnetic anisotropy is very low, 
about $6\times 10^{-6}$ of the exchange field, and no deviation from cubic 
symmetry was seen at $T_c$.

In our simulations\cite{TsaBunLan}, we used simple cubic lattices with
$12\le L\le 60$ at $T\le T_c=1.442929(77)|J|$. 
Numerical integrations of the coupled equations of motion were performed to a 
maximum time $t_{max} \le 1000|J|^{-1}$, using the algorithm based 
on 4th-order Suzuki-Trotter decompositions of exponential operators, with a 
time step $\delta t=0.2|J|^{-1}$ . 
As many as $7000$ initial configurations were used, although 
for large lattices this was reduced to as few as $400$.  
We were limited to the [100], [110] and [111] directions, i.e. 
${\bf q}=(q,0,0)$, $(q,q,0)$, $(q,q,q)$ and the equivalent momenta. 
For this model the order parameter is not conserved and the dynamic
structure factor cannot be separated into a longitudinal and a transverse
component. Henceforth we will use the term dynamic structure factor $S(q,\omega)$
and characteristic frequency $\bar\omega_m$ to refer to the average. 

\begin{figure}
\epsfxsize=3.0in
\vspace{-3cm}
\centerline{\epsfbox{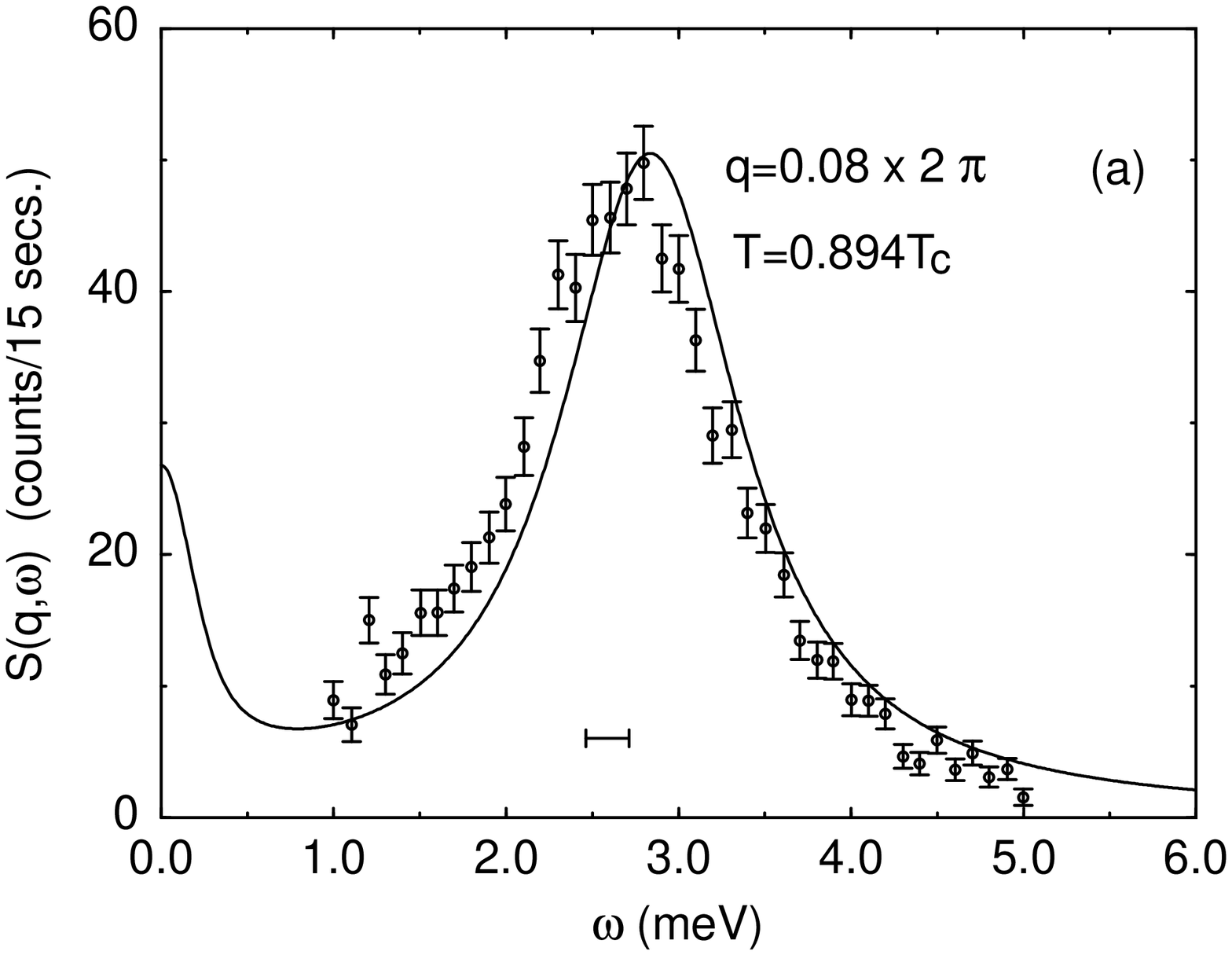}}
\epsfxsize=3.4in
\vspace{-5.0cm}
\centerline{\epsfbox{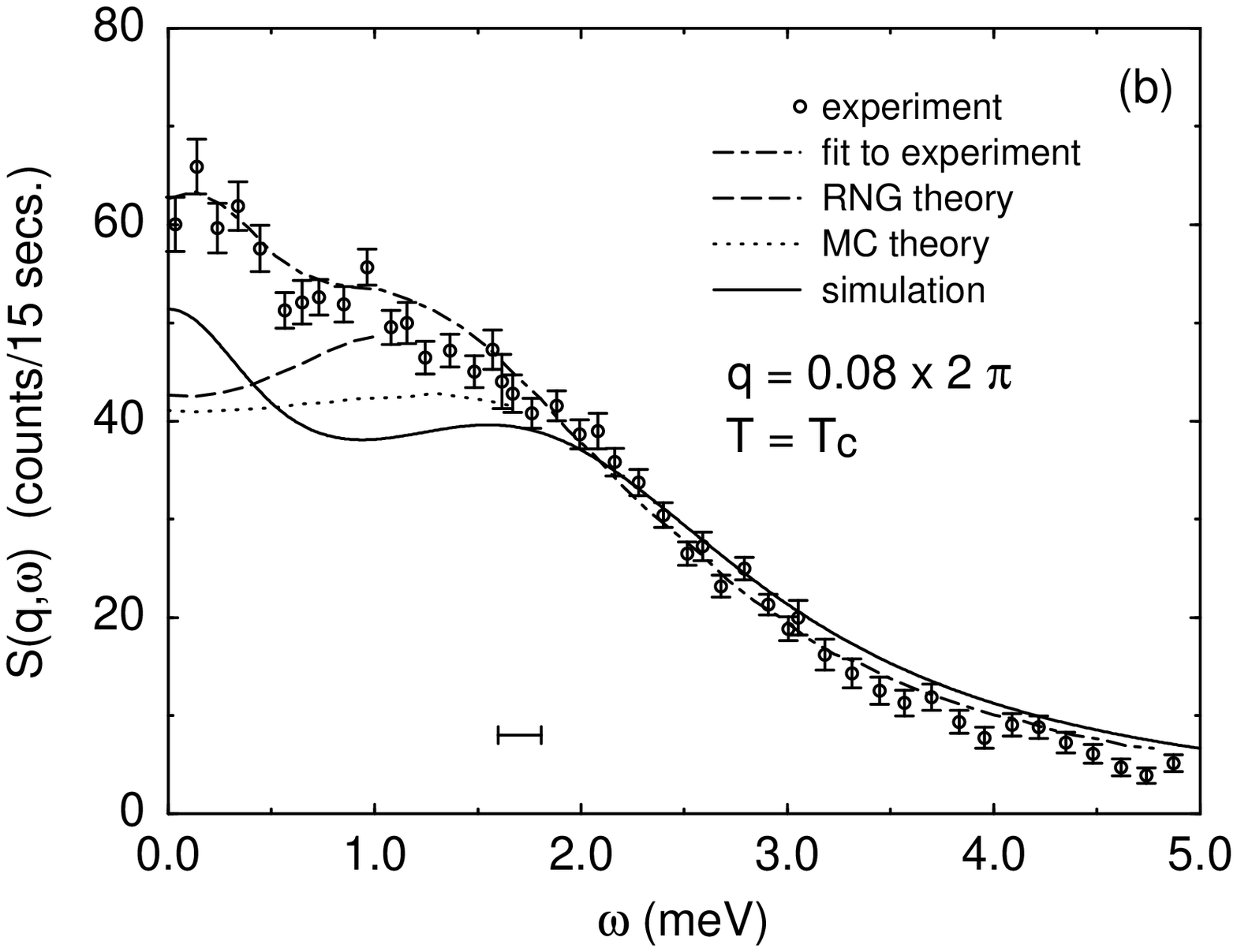}}
\vspace{-0.9cm}
\caption{Comparisons of lineshapes obtained from fits to
simulation data for $L=60$ (solid line) and the experiment (open circles) for
$q=2\pi(0.08)$ in the [111] 
direction, at (a) $T=0.894T_c$ and (b) $T=T_c$. 
The dot-dashed line in (b) is a fit to the experimental data 
which is compared to the predictions of renormalization group (RNG) 
and mode coupling (MC) theory. The horizontal line segment in each graph 
represents the $0.25$meV resolution in energy (full-width at half-maximum).}
\label{cexpfig}
\end{figure}

We compare our results with the recent neutron scattering data of Coldea 
{\it et al}.\cite{ColEtal} Although the compound RbMnF$_3$ is a quantum system, 
and our simulations are for a classical Hamiltonian, 
it has been shown that quantum Heisenberg systems with large spin values $(S\ge 2)$
behave as classical Heisenberg systems where the spins are vectors of magnitude
$\sqrt{S(S+1)}$ with the same interaction strength. Since our classical spins
were vectors of unit length, to preserve the Hamiltonian and the dynamics, a
normalization of the interaction strength and the frequency transfer are needed, 
i.e. $J=J^{exp}S(S+1)$ and $\omega^{exp}=J^{exp}\sqrt{S(S+1)}\; w/J$. For our
comparison, the experimental $T$- and $\omega$-dependent population factor was
removed from the experimental data, and the lineshapes from our simulation were 
convoluted
with the experimental Gaussian resolution function in energy. Fig.\ref{cexpfig}
shows the direct comparison of $S(q,\omega)$ in the [111] direction from 
simulation and experiment for $q=2\pi (0.08)$ [in this notation the Brillouin
zone edge in the [111] direction corresponds to $q=2\pi(0.25)$],
at $T=0.894T_c$ and at $T_c$. Below $T_c$, renormalization group theory 
\cite{MazNolFre} (RNG)
predicts a spin-wave peak and a central peak in the longitudinal component of $S(q,\omega)$; 
however, at $T_c$, both RNG\cite{FreMaz} and mode coupling\cite{CucLovTog} 
(MC) theory predict only the
presence of a spin-wave peak, while the experiment and the simulation find
a spin-wave peak and a central peak at $T=T_c$ as well (see Fig.\ref{cexpfig}(b)).
At low temperatures the central peak has very low intensity and the dominant structures are 
very narrow and sharp spin-wave peaks, from which accurate dispersion curves could be found. 
The dispersion curve for small $q$ changes from
a linear behavior at low $T$ to a power-law relation as $T\to T_c$. 
 
The dynamic critical exponent $z$ was extracted from the slope of a $\log(\bar\omega_m)$
vs $\log(L)$ plot (see Eq.(\ref{omegam})) corresponding to the [100] direction, using 
lattices in the asymptotic-size regime 
($L\ge 30$), and keeping $qL$ and $\delta_{\omega}L^z$ constant. For $\delta_{\omega}=0$,
we find $z=1.45(1)$ for $n=qL/(2\pi)=1$ and $z=1.42(1)$ for $n=2$. Using $\delta_{\omega}\neq 0$
requires an iterative procedure and the converged values that we obtained are 
$z=1.43(1)$ for $n=1$ and $z=1.42(1)$ for $n=2$. Hence, our final estimate is
$z=1.43(3)$, which is in agreement with the recent experimental value $z=1.43(4)$,
and slightly lower than the theoretical prediction $z=d/2=1.5$.
\begin{figure}
\vspace{-2.5cm}
\epsfxsize=3.1in
\centerline{\epsfbox{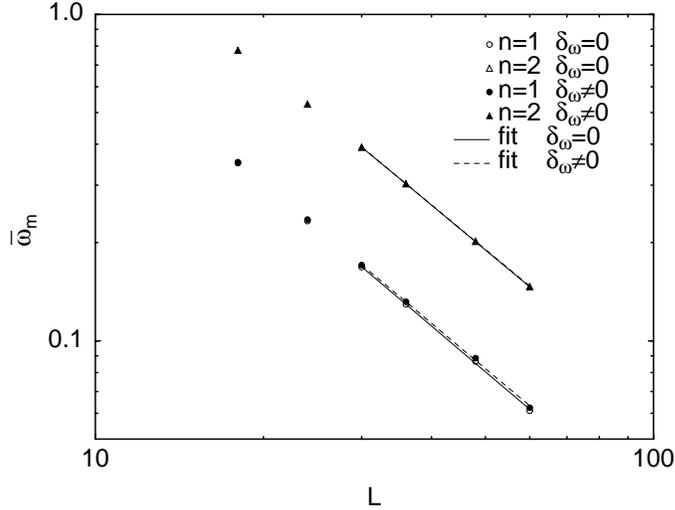}}
\vspace{-0.9cm}
\caption{ Finite-size scaling plot for $\bar\omega_m$ (with $qL=$const, 
$\delta_{\omega}L^z=$const) for the analysis with and without a resolution function. 
For the former case, the data used correspond to the converged values of $z$, for
$n=1,2$. The error bars were smaller than the symbol sizes.}
\label{fzs}
\end{figure}

\section{Summary}

We have shown how spin-dynamics techniques can be used to study critical and low-temperature 
magnetic excitations 
using simple classical spin models that have true dynamics, governed by equations of
motion. The solution of these equations is generally possible through the use 
of algorithms based on Suzuki-Trotter
decompositions of exponential operators and we compare their relative performance with each
other and with a predictor-corrector method. As examples of interesting physical systems,
we studied the two-dimensional XY model and the three-dimensional Heisenberg model.
We determined dynamic structure factors and through a finite-size scaling we estimated
the dynamic critical exponent of these systems. We have also made comparisons with
theoretical predictions and experimental data. 

\section*{Acknowledgements}

This work was partially supported by NSF grant No. DMR-9727714. 
Our simulations were carried out on the Cray T90 at the San Diego 
Supercomputing Center, and on an SGI Origin2000 
and IBM R6000 in the University of Georgia.


\begin{thebibliography}{99}

\bibitem{HohHal77}
 P.C. Hohenberg and B.I. Halperin, Rev. Mod. Phys. {\bf 49}, 435 (1977).
\bibitem{EvLan96}
 H.G. Evertz and D.P. Landau, Phys. Rev. B {\bf 54}, 12302 (1996).
\bibitem{ChenLan94}
 K. Chen and D.P. Landau, Phys. Rev. B {\bf 49}, 3266 (1994).
\bibitem{BunChenLan96}
 A. Bunker, K. Chen, and D.P. Landau, Phys. Rev. B {\bf 54}, 9259
 (1996).
\bibitem{GerLan90}
 R.W. Gerling and D.P. Landau, Phys. Rev. B {\bf 41}, 7139 (1990).
\bibitem{RapLan96}
 D.C. Rapaport and D.P. Landau, Phys. Rev. E {\bf 53}, 4696 (1996).
\bibitem{FraHuaLei} F. Frank, W. Huang, B. Leimkuhler, J. Comp. Phys. {\bf 133},
 160 (1997).
\bibitem{KreBunLan} M. Krech, A. Bunker, and D.P. Landau, Comput. Phys.
 Commun. {\bf 111}, 1 (1998).
\bibitem{NumAna81}
 R.L. Burden, J.D. Faires, and A.C. Reynolds, {\em Numerical Analysis}
 (Prindle, Weber \& Schmidt, Boston, 1981), p.205 and p.219.
\bibitem{SuzUme93}
 M. Suzuki and K. Umeno in {\em Computer Simulation Studies in
 Condensed Matter Physics VI}, edited by D.P. Landau, K.K. Mon, and
 H.B. Sch\"uttler (Springer, Berlin, 1993), p. 74.
\bibitem{KosTho73}
 J.M. Kosterlitz and D.J. Thouless, J. Phys. C {\bf 6}, 1181 (1973).
\bibitem{Vil74}
 J. Villain, J. Phys. (Paris) {\bf 35}, 27 (1974).
\bibitem{NelFis77}
 D.R. Nelson and D.S. Fisher, Phys. Rev. B {\bf 16}, 4945 (1977).
\bibitem{TsaBunLan}
 S.-H. Tsai, A. Bunker, and D.P. Landau, Phys. Rev. B, in press (see references therein).
\bibitem{ColEtal}
 R. Coldea, R.A. Cowley, T.G. Perring, D.F. McMorrow, and B. Roessli,
 Phys. Rev. B {\bf 57}, 5281 (1998).
\bibitem{MazNolFre}
 G.F. Mazenko, M.J. Nolan, and R. Freedman, Phys. Rev. B {\bf 18}, 2281 (1978).
\bibitem{FreMaz}
 R. Freedman and G.F. Mazenko, Phys. Rev. Lett. {\bf 34}, 1575 (1975);
 Phys. Rev. B {\bf 13}, 4967 (1976).
\bibitem{CucLovTog}
 A. Cuccoli, S.W. Lovesey, and V. Tognetti, J. Phys.: Condens. Matter {\bf 6},
 7553 (1994).

\end{thebibliography}
\end{document}